\documentclass[fleqn,10pt]{wlscirep}
\usepackage[utf8]{inputenc}
\usepackage[T1]{fontenc}
\usepackage{tikz,lipsum,lmodern}
\usepackage[most]{tcolorbox}
\usepackage{bm}
\title{The quantum Mpemba effects}

\author[1]{Filiberto Ares}
\author[1,2]{Pasquale Calabrese}
\author[3,4,5]{Sara Murciano}
\affil[1]{SISSA and INFN, via Bonomea 265, 34136 Trieste, Italy }
\affil[2]{International Centre for Theoretical Physics (ICTP), Strada Costiera 11, 34151 Trieste, Italy}
\affil[3]{Walter Burke Institute for Theoretical Physics, Caltech, Pasadena, CA 91125, USA}
\affil[4]{Department of Physics and IQIM, Caltech, Pasadena, CA 91125, USA}
\affil[5]{Universit\'e Paris-Saclay, CNRS, LPTMS, 91405, Orsay, France}

\begin{abstract}

The Mpemba effect, where a hotter system can equilibrate faster than a cooler one, has long been a subject of fascination in classical physics. 
In the past few years, significant theoretical and experimental progress has been made in understanding its occurrence in both classical and quantum systems. In this review, we provide a concise overview of the Mpemba effect in quantum systems, with a focus on both open and isolated dynamics which give rise to distinct manifestations of this anomalous non-equilibrium phenomenon. 
We discuss key theoretical frameworks, highlight experimental observations, and explore the fundamental mechanisms that give rise to anomalous relaxation behaviors. 
Particular attention is given to the role of quantum fluctuations, integrability, and symmetry in shaping equilibration pathways.  Finally, we outline open questions and future directions.

\end{abstract}
\begin{document}

\flushbottom
\maketitle

\thispagestyle{empty}


\section*{Introduction}

All physicists encounter the principles of thermodynamics during their undergraduate studies, learning that when the temperature of a macroscopic object is gradually changed, it remains in thermodynamic equilibrium. This ensures that its time-dependent state can be described by a well-defined temperature at each moment. Consequently, cooling a hot object is expected to occur incrementally, passing through all intermediate temperatures. Following this logic, a colder object should always reach equilibrium at a lower temperature faster than a hotter one.
However, the Mpemba effect challenges this deeply ingrained intuition. It reveals a surprising phenomenon: under certain conditions, a hot sample can cool faster than a colder one~\cite{mo-69}. This counterintuitive observation has intrigued and puzzled scientists for centuries. Historical accounts~\cite{aristotle, rbacon, fbacon, descartes, groves} suggest that the effect has been debated for millennia, and yet it continues to stir controversy~\cite{bl-16, bh-20} and remain a topic of active investigation within the scientific community~\cite{glcg-11, ahn-16, lvps-17, keller-18, hu-18, bj-19, sl-22, hr-22, tyr-23, gamlmms-24, lr-17, krhv-19, kb-20, bkc-21, gr-20, kcb-22, wv-22, wbv-23, bwv-23, santos-24, vvh-24,rev}.
Despite its long history, the Mpemba effect lacks a universally accepted explanation. 
Researchers have proposed various mechanisms, ranging from differences in thermal conductivity, evaporation rates, and initial energy distributions to more intricate considerations of nonequilibrium dynamics. 
Nevertheless, the effect remains a fascinating anomaly that defies simple thermodynamic reasoning, inviting further exploration to uncover its underlying principles.

Rather than delving into the ongoing debate surrounding the Mpemba effect in classical systems, we turn our focus to the quantum realm, exploring whether similar phenomena can arise there. As we will review, not only do quantum analogs of the Mpemba effect exist, but their manifestations are also less controversial, providing a clearer and more rigorous framework for understanding the phenomenon.
The first step is to clarify what we mean by the quantum Mpemba effect. In this context, we define it as an anomalous dynamical process where a system initially further from equilibrium relaxes faster than a system closer to equilibrium. This broad definition encompasses phenomena such as the cooling of a quantum system, as well as its heating, the latter often referred to as the inverse Mpemba effect.
Interestingly, even in quantum systems, such relaxation processes are typically governed by classical fluctuations rather than quantum ones. 
Beyond these cases, our definition also opens the door to a distinct class of phenomena involving the evolution of closed quantum systems. Specifically, consider a system prepared in a pure, nonequilibrium state and allowed to evolve under unitary dynamics—commonly referred to as a quantum quench. 
In this scenario, we ask whether it is possible for a system that begins further from equilibrium to relax faster than one that starts closer to it. Unlike the earlier examples, this type of relaxation is entirely driven by quantum fluctuations, offering a fundamentally different perspective on nonequilibrium behavior in quantum systems.
Through this lens, the quantum Mpemba effect becomes a unifying concept that spans both open and closed systems, connecting classical-like relaxation dynamics to the uniquely quantum processes that govern unitary evolution. This duality highlights the rich and multifaceted nature of nonequilibrium phenomena in the quantum world. 

In the following, we will explore both classes of the quantum Mpemba effect, examining their occurrence in natural and controlled quantum systems. These two classes—relaxations driven by classical-like fluctuations in open quantum systems and unitary dynamics driven by quantum fluctuations in closed systems—offer complementary perspectives on how the Mpemba effect manifests in the quantum realm.
A central challenge in analyzing the quantum Mpemba effect is determining how to quantify the distance of a system from equilibrium. This is a crucial step, as it enables meaningful comparisons between the relaxation dynamics of systems starting at different levels of disequilibrium. Without a robust measure of deviation from equilibrium, it would be impossible to characterize or verify the counterintuitive nature of the Mpemba effect.
To address this, we dedicate the first section of this work to defining and discussing metrics for quantifying a system's departure from equilibrium.  By establishing a clear framework for this measurement, we pave the way for a rigorous analysis of the quantum Mpemba effect across its various manifestations.

\section*{Quantifying the distance from equilibrium}

There are many reasonable candidates to quantify the separation of a system from equilibrium. In Box~1, we collect some of the most common choices in the literature to study the quantum Mpemba effect. One possibility is to directly take the density matrices that describe the state of the system (or a part of it) during the time evolution, $\rho(t)$, and the final equilibrium state, $\rho_{ss}$, and calculate one of the standard distances between matrices, such as the trace or the Frobenius (Hilbert-Schmidt) distances. Each distance has its pros and cons when studying the quantum Mpemba effect. For example, the Frobenius distance is the easiest to compute but it is not always monotonic in time in Markovian dynamics~\cite{ws-09}, unlike the trace distance. Although not symmetric as required for being a mathematical distance, the quantum relative entropy between $\rho(t)$ and $\rho_{ss}$ is also a measure of the distinguishability between two quantum states. Its classical analogue, the Kullback-Leibler divergence, is widely employed to investigate the Mpemba effect in classical Markovian systems~\cite{lr-17, vvh-24}. This relative entropy is closely related to the non-equilibrium free energy, which is also a suitable thermodynamic quantity to define and analyze the Mpemba effect in quantum open systems~\cite{goold}.

\begin{figure}[t]
\centering
\includegraphics[width=\linewidth]{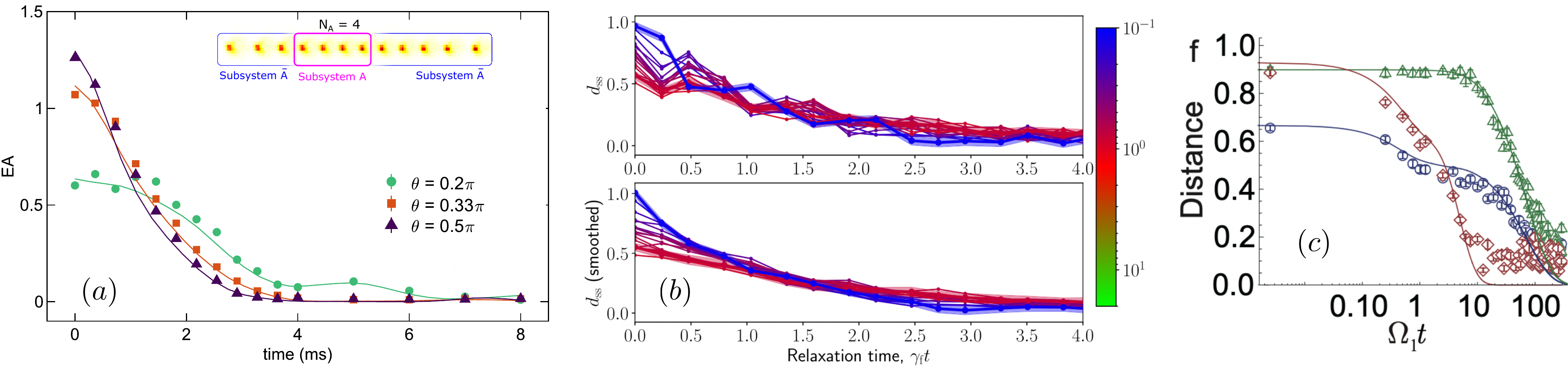}
\caption{\textit{Experimental observation of three quantum Mpemba effects.} Panel (a) corresponds to the experiment performed in Ref.~\citeonline{joshi-24}: a 12-site spin chain is quenched from several configurations, parameterized by $\theta$, which break to a different extent a $U(1)$ symmetry. The relaxation to equilibrium is monitored by studying the restoration of the symmetry in the 4-site subsystem $A$ using the R\'enyi-2 asymmetry. Panel (b) is the experiment in Ref.~\citeonline{shapira}, in which the inverse Mpemba effect is demonstrated: a single qubit is prepared at different initial temperatures (indicated by the color scale) and relaxes to the same steady state at a larger temperature. Panel (c) shows the experiment in Ref.~\citeonline{zhang2024}, where the strong Mpemba effect is observed: a single trapped ion is prepared in three different initial configurations and equilibrates following a Markovian dynamics to the same steady state. The initial state in red is obtained by applying a unitary rotation to the green initial state, which makes it relax exponentially faster than in the green and blue cases. In panels (b) and (c), it is plotted the trace distance between the time-evolved state and the final equilibrium state. Panel~(a) is reprinted from Ref.~\citeonline{joshi-24}. Panel~(b) is reprinted from Ref.~\citeonline{shapira}. Panel~(c) is reprinted from Ref.~\citeonline{zhang2024}.  }
\label{fig:experiments}
\end{figure}

\begin{figure}[t]
\centering
\includegraphics[width=0.65\linewidth]{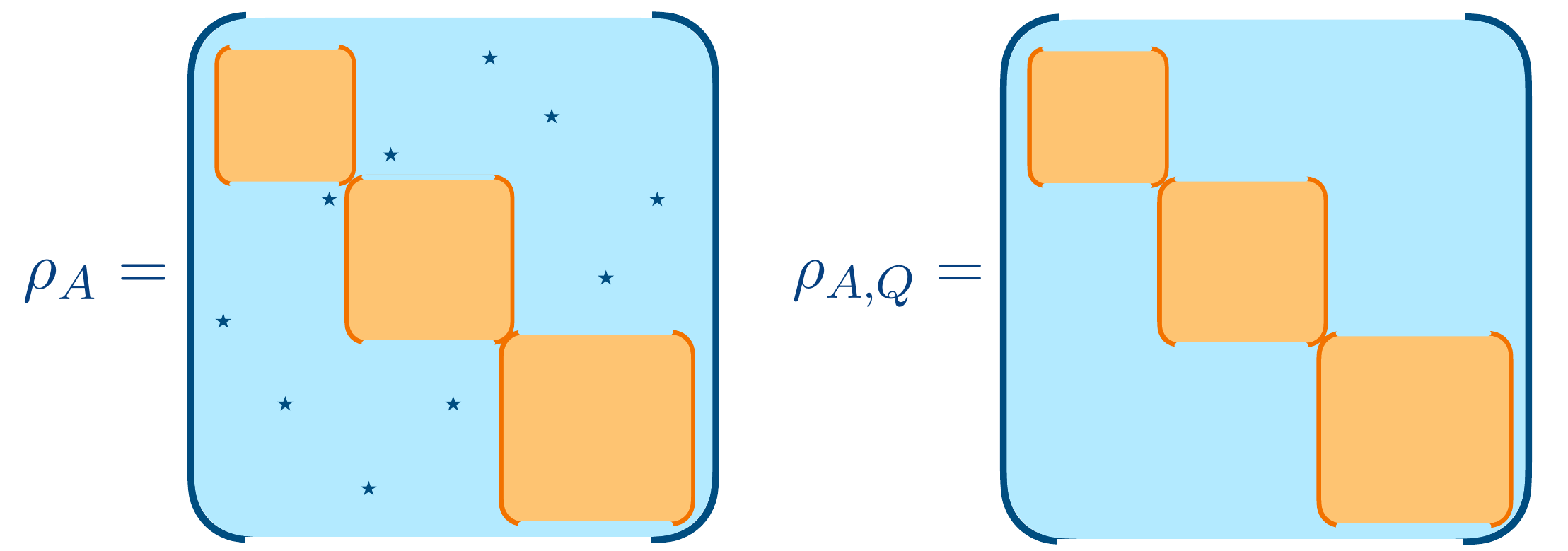}
\caption{\textit{Reduced density matrix $\rho_A$ and its symmetrization $\rho_{A, Q}$ that enter in the definition of the entanglement asymmetry}~\cite{amc-23}. We consider a $U(1)$ symmetry generated by a charge $Q$ that decomposes into the contributions of the subsystems $A$ and $\bar{A}$, $Q=Q_A+Q_{\bar A}$. If $\rho_A$ is not symmetric, i.e. $[\rho_A, Q_A]\neq 0$, then it is not block-diagonal in the eigenbasis of $Q_A$. By projecting over the eigenspaces of $Q_A$, we get the block-diagonal density matrix $\rho_{A, Q}$, which is symmetric, $[\rho_{A, Q}, Q_A]=0$. Each block corresponds to a symmetry sector with a fixed charge. The entanglement asymmetry quantifies how far $\rho_A$ is from being block diagonal. Figure reprinted from Ref.~\citeonline{amc-23}. }
\label{fig:rhoAQ}
\end{figure}

Another option is to indirectly monitor how fast the equilibrium is reached. For example, when the initial and final states are in different thermal phases, we can use an order parameter, such as the magnetization. In the quantum quench of a closed system, the determination of the final local equilibrium state is sometimes challenging. An alternative is employing a symmetry that is broken by the initial state and locally restored by the time evolution as a proxy for relaxation. One might also consider here an order parameter to detect symmetry breaking, since it indicates that the symmetry is broken when it does not vanish. However, the inverse is not always true, the order parameter may vanish when the symmetry is broken. Thus, a broken symmetry is not always captured by an order parameter. This ambiguity is resolved by using the entanglement asymmetry~\cite{amc-23}, whose definition finds its roots in the theory of entanglement in many-body quantum systems and coincides with quantities formerly studied in resource theory~\cite{vawj-08, gms-09, cg-19} and algebraic quantum field theories~\cite{chmp-20, chmp-21}. Typically, we consider a bipartite system $A\cup\bar{A}$, where $A$ is characterized by its reduced density matrix $\rho_A(t)$. To measure how much $\rho_A(t)$ breaks a symmetry, we construct a symmetrized version of $\rho_A(t)$, denoted as $\rho_{A, Q}(t)$ (we sketch in Fig.~\ref{fig:rhoAQ} the construction of $\rho_{A, Q}$ in the case of a $U(1)$ symmetry). Then the R\'enyi entanglement asymmetry is the difference between the R\'enyi entropies of $\rho_{A, Q}(t)$ and $\rho_A(t)$. The asymmetry quantifies the ‘distance’ between $\rho_A(t)$ and its symmetrized counterpart. Unlike the order parameter, it encodes all the information about the breaking of symmetry, including the non-local correlations, from the state itself through its dependence on the density matrix. It is non-negative and vanishes if and only if $\rho_A(t) = \rho_{A,Q}(t)$; that is, when the state of $A$ respects the symmetry. In the limit $n\to 1$, the entanglement asymmetry corresponds to the relative entropy between $\rho_A(t)$ and $\rho_{A, Q}(t)$. However, it is much easier to access it both theoretically and experimentally for integer values of $n$. The main disadvantage of the entanglement asymmetry is that it can only be used as a probe of the quantum Mpemba effect in those cases in which an initially broken symmetry is restored by the time evolution.

\begin{tcolorbox}[colback=blue!5!white,colframe=white]
 \textbf{Box 1 | Quantifying the distance from equilibrium}\\
\textbf{Trace distance}
\vspace{-0.8em}
\begin{itemize}
    \item \textbf{Definition}: 
$d_{\mathrm{Tr}}(\rho(t))=1/2\,\mathrm{Tr}\sqrt{A^{\dagger}A}$, where $A=\rho(t)-\rho_{ss}$, $\rho(t)$ is the time-evolved density matrix and $\rho_{ss}$ is the final equilibrium state
    \vspace{-0.7em}\item \textbf{Applications}: It is monotonic in time under Markovian dynamics. It has been used\cite{zhang2024,shapira} to probe experimentally the strong and inverse quantum Mpemba effect (see panels (b) and (c) of Fig.~\ref{fig:experiments})
\end{itemize}
 \vspace{-0.5em}
\noindent \textbf{Frobenius distance}
\vspace{-0.8em}
\begin{itemize}
    \item \textbf{Definition:}  
$d_{\mathrm{F}}(\rho(t))=\sqrt{\mathrm{Tr}\,A^{\dagger}A}$, where $A$ is again $A=\rho(t)-\rho_{ss}$, and it is easier to compute than the trace distance 
   \vspace{-0.7em}
   \item \textbf{Applications}: It has been used \cite{cll-21} to observe the Mpemba effect in Markovian systems. Unlike the trace distance, it is not monotonic in time in that case.
\end{itemize}
 \vspace{-0.5em}
\noindent\textbf{Quantum relative entropy }
\vspace{-0.8em}
\begin{itemize}
    \item \textbf{Definition:} 
$S(\rho(t)||\rho_{ss})=\mathrm{Tr}[\rho(t)(\log \rho(t)-\log \rho_{ss})]$
  \vspace{-0.7em} \item \textbf{Applications}: One application to explore the Mpemba effect can be found in Ref. \citeonline{Hayakawa0}. It is also closely related to the non-equilibrium free energy, used \cite{goold} to investigate the quantum Mpemba effect. 
\end{itemize}
\vspace{-0.5em}
\noindent\textbf{Order parameter}
\vspace{-0.8em}
\begin{itemize}
    \item \textbf{Definition:} Measurable quantity characterizing the phase of a system, often associated with symmetry-breaking phenomena
  \vspace{-0.7em} \item \textbf{Applications}: An example is the magnetization, used \cite{nf-19} to explore the quantum Mpemba effect. If it does not vanish, it signals the symmetry breaking, but the inverse is not true
\end{itemize}
\vspace{-0.5em}
\noindent\textbf{Entanglement asymmetry}
\vspace{-0.8em}
\begin{itemize}
    \item \textbf{Definition:} $\Delta S_A^{(n)}=S^{(n)}(\rho_{A,Q})-S^{(n)}(\rho_{A})$, where $\rho_A$ is the reduced density matrix, $\rho_{A,Q}$ is its symmetrized version, and $S^{(n)}(\rho)=\frac{1}{1-n}\log \mathrm{Tr}\rho^n$ is the R\'enyi entropy
  \vspace{-0.7em} 
  \item \textbf{Technicality}: To compute it, one introduces an auxiliary quantity called charged moment $Z_n(\boldsymbol{\alpha})={\rm Tr}\left(\prod_{j=1}^n \rho_A e^{i(\alpha_j-\alpha_{j+1})Q_A}\right)$, with $\boldsymbol{\alpha}=(\alpha_1,\dots, \alpha_n)$ and $Q_A$ is the charge generating the symmetry. In terms of it,
  \begin{equation}\label{eq:asymm_charged_moment}
  \Delta S_A^{(n)}=\frac{1}{1-n}\log\int_{-\pi}^{\pi}\frac{{\rm d}\alpha_1\cdots{\rm d}\alpha_n}{(2\pi)^n}\frac{Z_n(\boldsymbol{\alpha})}{Z_n(\boldsymbol{0})}
  \end{equation}
  \item \textbf{Applications}: If  an initially broken
symmetry is restored by the time evolution, it probes the quantum Mpemba effect \cite{amc-23} (see the panel (a) of Fig. \ref{fig:experiments})
\end{itemize}
\end{tcolorbox}

The behavior in time of some of the previous distances when the quantum Mpemba effect occurs is shown in Fig.~\ref{fig:experiments}. Generically, to study the Mpemba effect, the same quantum system is prepared in two different initial states described by the density matrices $\rho_1(0)$ and $\rho_2(0)$ and it is then evolved in time. If $d(\rho_1(0))>d(\rho_2(0))$, where $d(\rho)$ is one of the distances in Box~1, then $\rho_1(0)$ is farther from the final equilibrium state than $\rho_2(0)$. The Mpemba effect occurs when there is a time $t_M$ after which this inequality is inverted and $d(\rho_1(t))<d(\rho_2(t))$ for any $t>t_M$. This implies that $d(\rho_1(t))$ and $d(\rho_2(t))$ intersect at certain time, as we can see in Fig.~\ref{fig:experiments}. This crossing is the main signature of the occurrence of the Mpemba effect. The time evolution may be either unitary, which describes a closed quantum system and is dictated by the standard Schrodinger equation, or non-unitary, when the system is open and interacts with its environment. The Mpemba effect can happen in both situations, which we will discuss in what follows.





\section*{Mpemba effect in open quantum systems}
The most natural way to investigate whether Mpemba-like phenomena can manifest in quantum systems is to mimic the classical counterpart, specifically by implementing a temperature quench. 
In classical systems, the Mpemba effect can happen in scenarios such as supercooling~\cite{auerbach-95, ers-08}, where a substance is cooled below its freezing point without undergoing solidification.  In this scenario, two phases coexist: the liquid phase and the supercooled metastable phase. 
The Mpemba effect occurs when a system that is suddenly cooled down from a lower temperature remains trapped in the metastable phase for a longer period than one that starts at a higher temperature, before ultimately relaxing to the true equilibrium state. 
When supercooling is performed at very low temperatures, quantum effects may be relevant. 
This served as the initial motivation for investigating the Mpemba effect in the quantum realm, as explored in Ref.~\cite{nf-19}. That study examined a quantum Ising model coupled to a thermal bath, where the presence of a first-order quantum phase transition and metastable phases provided a natural setting for observing Mpemba-like behavior. 
After a quench of the temperature of the bath, the system follows a dissipative dynamics based on the Lindblad equation, a generalization of the Schrodinger equation that takes into account the interaction of the system with the environment. When the system is initially prepared at two different temperatures, the magnetization in both cases exhibits a significant transient period, indicating that the system remains trapped in a metastable phase. After some time, the magnetization abruptly drops to its equilibrium value, but this transition occurs earlier in the initially hotter system. This demonstrates that quantum effects do not spoil the Mpemba effect in the context of supercooling.

Beyond supercooling, a general theory of the Mpemba effect has been recently developed~\cite{lr-17, krhv-19} and experimentally checked~\cite{kb-20, bkc-21} for classical Markovian systems, in which the interaction with the thermal bath is modeled as a stochastic process.
It is then natural to wonder whether a Mpemba effect also exists in Markovian open quantum systems. The first paper addressing this question was Ref.~\citeonline{cll-21}. As in quantum supercooling, the time evolution of the state of these systems is governed by a Lindblad equation, instead of a Fokker-Planck equation as in the classical case~\cite{lr-17, kb-20}. Both equations are, nevertheless, linear in the state of the system, and, therefore, a generic time-evolved state, either classical or quantum, can be written as an infinite series of the eigenmodes of the corresponding equation. The relaxation of the system to the equilibrium is governed by the slowest-decaying eigenmode. Generic initial states have contributions from all the eigenmodes,
including the slowest one, and all of them are excited during the evolution. 
The key result~\cite{cll-21} is that, in Markovian quantum systems, one can always find a unitary rotation that applied to an arbitrary initial state completely suppresses the contribution coming from the slowest eigenmode. 
In that case, the system relaxes to the stationary state exponentially faster. This phenomenon has been dubbed the strong Mpemba effect in the classical Markovian systems~\cite{krhv-19, kb-20}, whose origin is the same as in the quantum version---the complete suppression of the slowest eigenmode. 
This theoretical proposal has been experimentally realized in Ref.  \citeonline{zhang2024}, see also Fig.~\ref{fig:experiments}~(c), which demonstrates how to engineer the rotation in the initial state that produces a strong Mpemba effect by applying efficient gate operations on a single trapped ion in its ground state. 


The choice of the unitary rotation which speeds up the relaxation to a steady state is the key tenet to observe the quantum Mpemba effect~\cite{cll-21,zhang2024}. However an exponential faster relaxation does not automatically imply the occurrence of the Mpemba effect. It is also necessary that the state that relaxes faster is initially farther from equilibrium. A way to obtain a unitary rotation that both speeds up the relaxation and yields the Mpemba effect is by exploiting the mathematical structure of the Lindbladian in thermalizing open systems. This was the focus of Ref.~\citeonline{goold}, which used special mathematical operations to describe how a quantum system thermalizes when interacting with its environment. Starting with a state where the density matrix has off-diagonal elements in a specific basis, and applying a unitary transformation, they showed that the thermalization process can exponentially speed up, while also increasing the free energy of the initial state. By using the non-equilibrium free energy to detect the quantum Mpemba effect, the unitary transformation leads to a crossing between the free energy of the original and transformed states, where the quantum Mpemba effect occurs.  A similar conclusion can be derived by probing this phenomenon via different quantities, for
instance by connecting quantum stochastic thermodynamics to information geometry~\cite{goold2}.

So far, we have shown that a temperature quench or specific engineering of the initial state in open systems are possible ways to observe the quantum Mpemba effect. However, this phenomenon does not always occur because of the contribution coming from the slowest relaxation mode. For instance, studying a quantum dot connected to two reservoirs, the authors of Ref.~\citeonline{Hayakawa0} found that it arises as a combined effect of the remaining relaxation modes.
Starting from two different initial conditions, after a quench to the same temperature and chemical potential, they observed the quantum Mpemba effect in the time evolution of the density matrix elements, together with different types of Mpemba phenomena, including the mixed Mpemba effect, where one initial temperature is lower than the steady-state value, and the inverse Mpemba effect, where
both initial temperatures are below the steady-state temperature. 
The quantum Mpemba effect has also been explored in non-Hermitian systems~\cite{Hayakawa1}, revealing the possibility of multiple crossings in various observables, such as ground state probability, energy, entropy, temperature, and the distance from the steady state. 

The inverse Mpemba effect has been experimentally
observed in a single trapped ion qubit~\cite{shapira} (this reverse effect has also been predicted~\cite{gr-20} and experimentally observed at classical level~\cite{kcb-22}). 
The qubit, which follows a unitary evolution, is coupled to a Markovian thermal bath. This causes the decoherence of the qubit and its eventual relaxation to a stationary state. Then the qubit is quenched by suddenly increasing the temperature of the bath. In this case, the trace distance serves as a probe of the Mpemba effect, see Fig.~\ref{fig:experiments}~(b). In sufficiently coherent qubits, this experimental setup also shows that a cold qubit can heat exponentially faster, which is the strong version of the inverse Mpemba effect.

Over the past year, numerous studies have explored the Mpemba effect in a wide range of open Markovian quantum systems, examining its occurrence under diverse conditions~\cite{manikandan-21, kcl-22, bh-22, ias-23, zhou-23, ww-24, liu-24, longhi-24, longhi-24-2, befb-24, ew-24, wsw-24, longhi-24-3,fs-24, qww-24, dong-24, mcblg-24, kcm-24, gsm-24, znez-25}. For example, Ref.~\citeonline{nava-24} considers the case of models coupled to multiple reservoirs and discusses the possibility of observing the Mpemba effect using electron currents, easier to access than distances between density matrices. Some of these papers include their own methods, different to those we have discussed here, to construct initial states that show super-accelerated relaxation. 

One can further extend the playground and consider a non-Markovian dynamics~\cite{strachan-24, wwbw-24}. The introduction of a finite memory time yields new versions of the effect, in addition to the ones also found in the Markovian case. In particular, one can prepare an initial state that reaches the steady state within the memory time, giving the fastest possible relaxation in this kind of systems~\cite{strachan-24}. This has been dubbed extreme quantum Mpemba effect.

\section*{Quantum Mpemba effect in isolated systems}
In the previous section, we have seen that the Mpemba effect can occur in open quantum systems in scenarios that usually mimic those considered in the study of the phenomenon at the classical level. 
In this section, we explore a quantum version of the Mpemba effect in isolated quantum systems after a global quantum quench. In this case, the system is prepared in a pure, non-equilibrium state and it evolves with the Schrodinger equation. Unlike in the previous situations, there is no interaction with an environment and the relaxation processes are fully driven by quantum fluctuations. Since the evolution is unitary, the system never globally equilibrates to a stationary state. Instead, we have to focus on a subsystem $A$. For generic large quantum systems, the complementary subsystem $\bar{A}$ acts as a bath and
the reduced density matrix $\rho_A$ is expected to asymptotically relax to a (generalized) Gibbs ensemble under unitary evolution~\cite{deutsch-91, srednicki-94, rdyo-07, rdo-08}. 
Similarly to the open case, one might take the distance between the time-evolved state, $\rho_A(t)$, and the stationary state, $\rho_{A,ss}$, as we will illustrate in the following section. Alternatively, a simpler approach is to investigate the local restoration of a symmetry that is broken by the initial state but preserved by the time evolution.
In one dimensional systems, the restoration of the symmetry can be understood with the Mermin-Wagner theorem.  
According to it, continuous symmetries cannot be spontaneously broken at finite temperature in one dimension. In the local equilibrium state after the quench, the finite energy density of the initial state acts as an effective temperature, typically leading to symmetry restoration, except in a few cases due to peculiar conservation laws~\cite{amvc-23}. In terms of symmetries, the quantum Mpemba effect happens when the more the symmetry is broken, the faster it is restored. The main tool that we use to probe the quantum Mpemba effect in this case is the entanglement asymmetry defined in Box~1.

\subsection*{Experiment}
Although the quantum Mpemba effect through symmetry restoration was first theoretically predicted \cite{amc-23}, it was experimentally observed just a few months later in a trapped-ion quantum simulator \cite{joshi-24}. The setup is the following: a system of $12$ ions that simulates a spin-$1/2$ chain is initialized into a product state with all the spins aligned along the $z$-direction, and then they are tilted away this axis by an angle $\theta$, breaking the $U(1)$ symmetry associated with the rotational invariance about the $z$-axis. This state then evolves with a long-range interacting Hamiltonian that possesses such $U(1)$ symmetry. The tilting angle $\theta$ tunes the extent to which the symmetry is initially broken. By selecting initial states with varying degrees of symmetry breaking, we observe a clear crossing in the entanglement asymmetry, see Fig.~\ref{fig:experiments}~(a), for any choice of a subsystem $A$ containing 4 ions. 
This behavior confirms that states with greater symmetry breaking are able to restore the symmetry more rapidly, also in the presence of realistic physical conditions such as interactions, global dephasing, decoherence, and weak disorder, as occurs in the experiments performed. 
One specific scenario examined involves allowing the system to evolve solely under the environmental noise of the laboratory, without any Hamiltonian unitary evolution. 
In this case, the symmetry is also restored for any tilting angle, but the entanglement asymmetries do not cross. The absence of crossings in this case highlights the role of quantum fluctuations in the occurrence of the Mpemba effect in quantum quenches.
But the entanglement asymmetry is not the only way to detect the quantum Mpemba effect. The Frobenius distance between the time-evolved state of a subsystem and its expected stationary state was also measured in this experimental quench protocol, showing the same qualitative behavior as the entanglement asymmetry in Fig.~\ref{fig:experiments}~(a). This investigation not only validates the quantum Mpemba effect, but it also probes the local thermalization of a non-integrable system to the theoretical stationary state. 

\subsection*{Microscopic origin in integrable systems}

The natural question that arises at this moment is why this phenomenon occurs. It is important to note that, as happens in the classical case, the quantum Mpemba effect is not always observable, as it is shown in the left panel of Fig. \ref{fig:asymmetry}: while a crossing is clearly visible between the blue and the brown or green lines depicting the dynamics of the entanglement asymmetry, it does not occur between the red and the other curves. Therefore, an essential point is to determine the conditions under which it takes place. A complete answer to these inquiries has so far been given in the case of one-dimensional many-body quantum systems that are integrable. For these cases, a microscopic mechanism and a set of necessary and sufficient conditions, based on the properties of the initial states, have been identified \cite{rylands-24}. 

Integrable quantum systems are a very special family of models characterized by possessing an infinite number of conserved quantities. The constraints imposed by their conservation laws make the non-equilibrium dynamics of these systems radically different from the chaotic (non-integrable) ones and largely facilitate their theoretical study. A crucial consequence of integrability is the existence of stable excitations or quasiparticles in terms of which we can effectively explain many of their equilibrium and non-equilibrium
properties, in particular the quantum Mpemba effect.

\begin{figure}[t]
\centering
\includegraphics[width=0.6\linewidth]{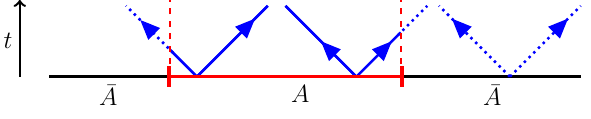}
\caption{\textit{Sketch of the quasiparticle picture for the entanglement asymmetry.} The quantum quench of a one-dimensional integrable system generates pairs of entangled quasiparticles that propagate ballistically with opposite momentum. Each pair contributes to the charged moment $Z_n(\boldsymbol{\alpha})$, from which the entanglement asymmetry can be calculated using Eq.~\eqref{eq:asymm_charged_moment}. When specializing to the quotient $Z_n(\boldsymbol{\alpha})/Z_n(\boldsymbol{0})$, only the pairs with both excitations inside the subsystem $A$ contribute (solid blue arrows). When one of the excitations leaves $A$, the pair stops contributing (dashed blue arrows). }
\label{fig:quasiparticle}
\end{figure}

In a quantum quench as those previously described, the initial configuration is a superposition of the eigenstates of the post-quench Hamiltonian. In an integrable model, these eigenstates are constituted by multiple excitations and, therefore, the initial state acts as a source of quasiparticles. After the quench, pairs of these quasiparticles are homogeneously emitted at each point of the system and propagate in opposite directions at a certain velocity. The pairs produced at different spatial points are uncorrelated while those quasiparticles emerging at the same point are entangled. These pairs of entangled quasiparticles are responsible for spreading entanglement and correlations through the system as they propagate far apart. This is the basis of the \textit{quasiparticle picture}~\cite{cc-05, ac-17, ac-18}, which has allowed to understand the time evolution of entanglement entropy and other observables after quenches in integrable systems. When a symmetry is broken in the initial state, the pairs of entangled quasiparticles carry correlations that do not respect such symmetry. Within this framework, the asymmetry measures the contribution to the breaking of the symmetry of the pairs that are inside the subsystem $A$, see Fig.~\ref{fig:quasiparticle}. The key 
point is that, when one of the excitations forming an entangled pair leaves the subsystem, then that pair stops contributing to breaking the symmetry in $A$ and, consequently, the asymmetry decreases. In an infinite system, the number of complete pairs in the subsystem $A$ tends to zero as they move far apart. The symmetry is then restored in the subsystem and the asymmetry goes to zero. This picture not only provides a microscopic mechanism for the restoration of the symmetry but also for the quantum Mpemba effect. If we consider two initial states, the symmetry is broken more in the one that contains a larger number of quasiparticle pairs that contribute more to symmetry breaking. However, after the quench, we have to take into account that each pair propagates at different velocity. This means that the symmetry is restored earlier if the pairs that break it more are also the fastest ones, as they take less time to abandon the subsystem. Therefore, the quantum Mpemba effect occurs when, in the more asymmetric initial state, the quasiparticles that transport the correlations that contribute more to symmetry breaking are faster than in the other initial state. This is the main result of Ref. \citeonline{rylands-24}, where the precise conditions for this to occur are derived. They only require the knowledge of the density of occupied quasiparticle modes in the initial state and their velocities after the quench. These two ingredients can be in principle determined within the Bethe ansatz framework for generic integrable systems.

\subsection*{Further results in integrable systems}

\begin{figure}[t]
\centering
\includegraphics[width=\linewidth]{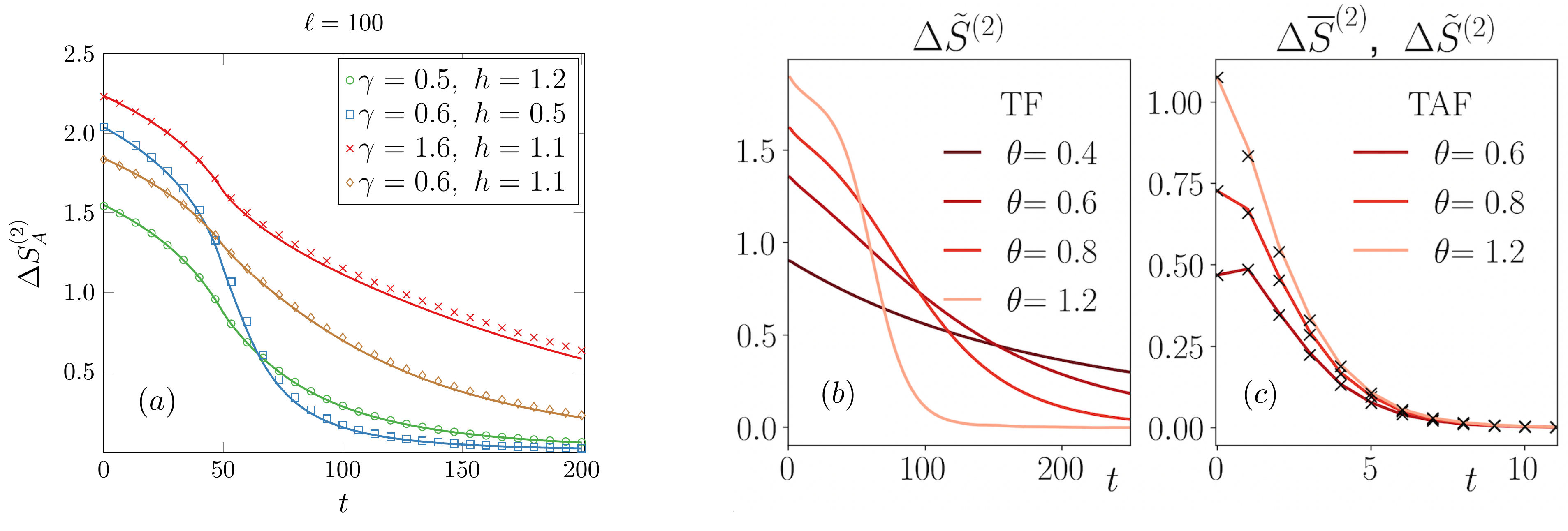}
\caption{\textit{Time evolution of the entanglement asymmetry in different systems.} Panel $a$ corresponds to Ref.~\citeonline{makc-23}, where the dynamics of the asymmetry is studied in a subsystem of $\ell=100$ sites of an infinite fermionic tight-binding chain from different initial states parameterized by $h$ and $\gamma$. Panel $b$ ($c$) shows the results obtained in Ref.~\citeonline{turkeshi-24} for the asymmetry of a tilted (anti)ferromagnetic state (with tilting angle $\theta$) that evolves in a charge-preserving random circuit. In this case, the subsystem size is 16 qubits and total system size is 512. Panel~(a) is reprinted from Ref.~\citeonline{makc-23}. Panels~(b) and~(c) are reprinted from Ref.~\citeonline{turkeshi-24}. }
\label{fig:asymmetry}
\end{figure}

The previous  quasiparticle framework has been applied to analyze the quantum Mpemba effect in several of the most paradigmatic integrable models: the fermionic tight-binding chain\cite{makc-23} (see panel ($a$) of Fig. \ref{fig:asymmetry}), where the quasiparticles can be identified as the Cooper pairs responsible for breaking the particle-number symmetry, in the Lieb-Liniger model, the rule 54 cellular automaton \cite{Klobas_2024} or the XXZ $1/2$-spin chain \cite{rylands2024}. Since the Mpemba physics is governed by the high-energy eigenstates, the groundstate phase of the post-quench Hamiltonian is in principle not relevant. However, in the XXZ spin chain, the occurrence of the phenomenon mimics the zero-temperature phase diagram of the model. Another surprising property observed in the fermionic tight-binding model is that the entanglement asymmetries of two different initial states can exhibit multiple crossings \cite{carc-24}. This result highlights that the existence of the quantum Mpemba effect cannot be inferred from the dynamics of the system at short times. The entanglement asymmetries usually cross at early times, but there may be other crossings at later times that spoil the Mpemba effect and lead to a normal relaxation. The quantum Mpemba effect has also been investigated through the restoration of discrete symmetries~\cite{fac-24}.

The previous studies restrict to one-dimensional systems, 
but does the quantum Mpemba effect survive in higher dimensions?
A first answer has been given\cite{yac-24} for a system of free fermions
on an infinite cylinder taking as subsystem a strip in the periodic direction. In this case, the quantum Mpemba effect is very sensitive to the length of this extra dimension. By varying it, the phenomenon can be enhanced or spoiled, depending on the choice of initial states.
This phenomenology can also be explained by extending the quasiparticle picture described above for one-dimensional systems. 
The quantum Mpemba effect has also been analyzed in a similar two-dimensional setup but quenching superfluid to free bosons \cite{yca-24}. In the superfluid phase, the particle number symmetry is spontaneously broken but it is not dynamically restored as the Bose-Einstein condensate remains after the quench. Thus this symmetry cannot be employed as a probe of the quantum Mpemba effect. Instead, the effect is studied considering the quantum fidelity between the time-evolved and the final stationary states. Remarkably, the 
appearance of the quantum Mpemba effect depends on the effective theory that we employ to describe the initial superfluid state. 

A quasiparticle picture is still valid in quantum quenches of some open quantum systems, such as free fermions with weak gain/loss dissipation~\cite{cma-24,avm-24}. Leveraging this tool, the quantum Mpemba effect can be observed for both generic ground states of short-range quadratic fermionic Hamiltonians, as well as for the tilted antiferromagnetic case, as in the many-body localized setup. The dissipation modifies the contribution of the excitations to the entanglement asymmetry, and it also speeds up the symmetry restoration. A remarkable feature of open systems is that the stationary state does not depend on the initial
condition of the quench: This scenario makes the dissipative setup more similar to the one
considered in the classical Mpemba effect\cite{kb-20}, where the relaxation of different initial states to a common equilibrium state is studied.
The quantum Mpemba effect has been studied via the asymmetry also in other setups, including conformal field theories \cite{benini2024} and periodically driven systems \cite{bds-24}.

\subsection*{Non-integrable systems}

The results above have shown that the mechanism behind the quantum Mpemba effect is well understood in integrable systems. Recent studies \cite{turkeshi-24,lzyz-24, krb-24, ylz-25, amcp-25} have also observed this phenomenon in chaotic quantum dynamics, expanding on the cases discussed so far. The generic unitary evolution is implemented through random circuits with $U(1)$ global symmetry. This analysis shows that, in chaotic systems, the effect also depends on the choice of the initial state: tilted ferromagnetic states can reveal the effect after a short time, whereas tilted antiferromagnetic states do not show it, as we can observe in the $(b)$ and $(c)$ panel of Fig. \ref{fig:asymmetry}, respectively. This behavior contrasts with many-body localized systems, where the quantum Mpemba effect appears for any initial tilted state, suggesting a different mechanism from that of chaotic systems~\cite{lzyzy-24-2}. Neither chaotic nor disordered systems admit a description in terms of quasiparticles, so the criterion described above cannot be applied. However, Ref.~\citeonline{turkeshi-24} provides an explanation for this phenomenon by exploiting the connection between the asymmetry and operator spreading. By relating the evolution of the asymmetry to the dynamics of non-conserved quantities, they understand why the quantum Mpemba effect does (or does not) occur when the quench begins from a (anti)ferromagnetic state.

Using the entanglement asymmetry, the quantum Mpemba effect has also been observed in dual-unitary quantum circuits with an arbitrary number of $U(1)$ conserved charges~\cite{foligno-24}. In these systems, the dynamics remains unitary upon exchanging the roles of space and time, which allows one to exactly calculate many observables that otherwise are very hard to compute. Despite possessing an arbitrary number of conserved charges, the dual-unitary circuits introduced in~\cite{foligno-24} show chaotic dynamics within each charge sector. Surprisingly, in quenches from certain states, they display a two-step relaxation. Entanglement typically grows linearly in time and eventually reaches a stationary value. Here instead it presents two different regimes before saturating, in which it linearly grows but at different rates. This two-step relaxation makes possible the faster restoration of a symmetry in the more asymmetric initial states.
Still, the precise conditions under which this two-step relaxation occurs remain unknown.

\section*{Outlook}

We have reviewed how, over the past few years, numerous theoretical and experimental advancements have significantly deepened our understanding of the quantum Mpemba effects across a wide range of physical settings. These studies have explored both open and closed systems, as well as few-body and many-body scenarios, uncovering intriguing mechanisms underlying the anomalous thermal relaxation phenomenon. However, despite this progress, many fundamental questions remain open, offering exciting opportunities for further investigation.

For open quantum systems, a compelling question is whether insights gained from the study of the quantum Mpemba effect can shed new light on its classical counterpart. Understanding the interplay between classical and quantum relaxation mechanisms could provide a broader theoretical framework for anomalous thermalization processes.

In the context of isolated many-body systems, one of the most pressing open questions concerns the general conditions under which the quantum Mpemba effect arises in ergodic Hamiltonian dynamics. While substantial progress has been made in understanding this effect in integrable models \cite{} and quantum circuits \cite{}, a unifying explanation for generic non-integrable many-body systems is still lacking. Investigating the role of chaos, eigenstate thermalization, and the structure of initial states in facilitating anomalous relaxation dynamics is a crucial step in addressing this question.

A particularly intriguing and unexplored avenue of research is the fate of the Mpemba effect in monitored quantum systems. This phenomenon can be studied in both Hamiltonian~\cite{kmr-23, fpsbn-23, ppgm-23} and circuit-based frameworks~\cite{srn-19, lcf-18, agrawal-22, cnps-19}, where the interplay between the Mpemba effect and measurement-induced entanglement transitions \cite{} could give rise to novel and unexpected phenomena. The impact of continuous monitoring on relaxation dynamics and the potential emergence of new universality classes in thermalization remain largely uncharted territories.

Another important direction concerns the effects of inhomogeneities and impurities on the quantum Mpemba effect. Disorder and spatial variations in system parameters can significantly alter thermalization pathways, potentially leading to novel manifestations of the effect. Investigating these scenarios may reveal new insights into non-equilibrium dynamics and anomalous heat flow.

Beyond fundamental theoretical questions, the Mpemba effect holds significant promise for practical applications, particularly in non-adiabatic state preparation. Traditional methods for preparing thermal states typically rely on slow, adiabatic cooling, which may not always be feasible in complex or rapidly evolving quantum systems. The quantum Mpemba effect offers an alternative, non-adiabatic pathway for preparing desired thermal states, particularly effective when starting far from equilibrium. This perspective opens up exciting possibilities in quantum control, quantum simulation, and quantum information processing, where efficient thermalization protocols could have direct technological implications.

In summary, while remarkable strides have been made in understanding the Mpemba effect in various physical settings, a wealth of open questions remains. From fundamental theoretical inquiries to potential technological applications, the continued exploration of this phenomenon promises to yield fascinating discoveries in the years to come.



\bibliography{sample}

\section*{Acknowledgements}
We are deeply grateful to the many physicists who have collaborated with us over the past few years on the study of the quantum Mpemba effect. 
In particular, we would like to thank Vincenzo Alba,  Bruno Bertini, Rainer Blatt, Fabio Caceffo, Konstantinos Chalas, Andrea De Luca, Florent Ferro, Alessandro Foligno, Johannes Franke, Lata Kh. Joshi, Manoj Joshi, Israel Klich, Katja Klobas, Florian Kranzl, Lorenzo Piroli, Aniket Rath, Christian Roos, Colin Rylands, Xhek Turkeshi, Benoit Vermersch, Eric Vernier, Vittorio Vitale, Shion Yamashika, and Peter Zoller. 
PC and FA acknowledge support from ERC under Consolidator Grant number 771536 (NEMO) and from European Union - NextGenerationEU, in the framework of the PRIN Project HIGHEST no. 2022SJCKAH\_002.
SM thanks the support from the Walter Burke Institute for Theoretical Physics and the Institute for Quantum Information and Matter at Caltech.

\section*{Author contributions}
The authors contributed equally to the writing of the manuscript.

\section*{Competing interests}

The authors declare no competing interests.

\end{document}